# Anomaly Detection in Unsupervised Surveillance Setting Using Ensemble of Multimodal Data with Adversarial Defense

Sayeed Shafayet Chowdhury
Purdue University

Kaji Mejbaul Islam
Ahsanullah University of Science and Technology

Rouhan Noor
Ahsanullah University of Science and Technology

*Abstract*— Autonomous aerial surveillance using drone feed is an interesting and challenging research domain. To ensure safety from intruders and potential objects posing threats to the zone being protected, it is crucial to be able to distinguish between normal and abnormal states in real-time. Additionally, we also need to consider any device malfunction. However, the inherent uncertainty embedded within the type and level of abnormality makes supervised techniques less suitable since the adversary may present a unique anomaly for intrusion. As a result, an unsupervised method for anomaly detection is preferable taking the unpredictable nature of attacks into account. Again in our case, the autonomous drone provides heterogeneous data streams consisting of images and other analog or digital sensor data, all of which can play a role in anomaly detection if they are ensembled synergistically. To that end, an ensemble detection mechanism is proposed here which estimates the degree of abnormality of analyzing the real-time image and IMU (Inertial Measurement Unit) sensor data in an unsupervised manner. First, we have implemented a Convolutional Neural Network (CNN) regression block, named AngleNet to estimate the angle between a reference image and current test image, which provides us with a measure of the anomaly of the device. Moreover, the IMU data are used in autoencoders to predict abnormality. Finally, the results from these two pipelines are ensembled to estimate the final degree of abnormality. Furthermore, we have applied adversarial attack to test the robustness and security of the proposed approach and integrated defense mechanism. The proposed method performs satisfactorily on the IEEE SP Cup-2020 dataset with an accuracy of 97.8%. Additionally, we have also tested this approach on an in-house dataset to validate its robustness.

## I. INTRODUCTION

Machine learning (ML) techniques have had far reaching impacts in diverse arenas ranging from image classification [1], assistive technologies [2-6], sensing [7-8] to heart rate measurement [9-10]. With the proliferation of automation, security has become an important issue for these technologies to detect and control malevolent activities. In this respect, autonomous surveillance is of particular interest due to its capability of continuous monitoring. In order to ensure safety of the perimeter that is under consideration, the autonomous agent must detect abnormalities using images and other available sensory data [11]. However, designing an effective system for such online detection tasks in resource-constrained autonomous bodies is challenging. Firstly, the type of anomaly might vary from previous scenarios encountered as the adversary has the degree of freedom for coming up with newer anomalies. Moreover, the moving drone has to perform detection in real-time with low latency and high accuracy since too many false positives will be expensive and false negatives might cause security breaches. Keeping these in mind, incorporating deep learning methods for anomaly detection seems attractive. Although supervised models give very high accuracy in general for detection purposes, we require a-priori information about all possible attack cases for these models to succeed, which may be an impractical assumption considering the possibility of novel abnormalities. Moreover, no matter how well the pipeline is designed, an attacker can fool the system using various adversarial attacks on the ML models [12]. So, for an effective and secure anomaly detection system, the defender must take into account the potential of such adversaries and take precautionary measures.

There have been extensive studies on anomaly detection using autonomous systems leveraging a wide variety of techniques. Campo *et al.* [13] proposed a Gaussian process (GP) regression method to detect abnormal motions in real vehicle situations based on trajectory data. A novel method based on internal cross-correlation parameters of the vehicle with Dynamic Bayesian Network (DBN) was implemented to determine the abnormal behavior in [14]. The authors in [15] utilized a method of selecting an appropriate network size for detecting abnormalities in multisensory data fed from a semiautonomous vehicle. Many of the previous works mostly attempt to learn private layer (PL) self-awareness models based on a high level of supervision [16-17]. On the contrary, Ravanbakhsh *et al.* [18] proposed a dynamic incremental self-awareness (SA) model which learns through experiences in a hierarchical manner, growing in complexity from simple to more structured cases. They used cross-modal Generative Adversarial Networks (GAN) to make usage of image data. Again, Baydoun *et al.* [19] also adopted a similar approach of training a set of GANs in a semi-supervised fashion to perform multi-sensor anomaly detection. It was implemented for moving cognitive agents using both external and private first-person visual observations to characterize agents' motion in a given environment. Notably, the only focus of most of these works lies in environmental anomaly detection. While this is an important and demanding task itself, another pertinent issue is the operating condition of the surveillance agent. Even if the detection system operates faithfully, the whole pipeline will be hampered greatly if the data collecting device is suffering from anomaly. However, the monitoring of device anomaly has usually not been taken into consideration in the literature. Recognizing this gap, the IEEE SP CUP 2020 competition [20] pivots on autonomous device anomaly detection in a surveillance setting and this is the problem we try to focus on in this paper. Another pivotal issue of modern ML based system's vulnerability is adversarial attacks on the learnt models. Interestingly, none of the previous works consider the failure cases that may arise due to such attack cases. Without a thorough investigation of the resistance of these models to some of these adversarial perturbations, it is unfair to claim their success against malicious anomalies.

To address the above mentioned critical issues in autonomous anomaly detection, in this paper, we introduce a modification of Siamese Network [21] consisting of Convolutional Neural Network (CNN) based regression architecture namely AngleNet which estimates angular displacement between two images. Additionally, our proposed method uses an autoencoder based anomaly detection system from the Inertial Measurement Unit (IMU) data available from the drone. Finally, an ensemble mechanism is utilized to estimate the degree of abnormality using predictions both from the given image and IMU data samples at a particular timestamp.

The main contributions of this paper are:

1. We have implemented an ensemble of multimodal data (image and sensory motion data) for accurate, low-latency device anomaly detection in real-time for autonomous surveillance.

2. To ensure an effective and secure anomaly detection system, we have taken into account the potential of adversarial attacks and integrated adversarial training as a defense mechanism to further strengthen this system.

The organization of the rest of the paper is as follows. Section II explains the details of the problem description. Section III describes the proposed method where the AngleNet and the rest of the clustering methods are explained in detail. Section IV presents the experimental results and comparison and finally, the work is concluded with section V.

II. PROBLEM AND DATASET DESCRIPTION

In SP Cup 2020 [20] competition dataset, Rosbag files were provided which contained data from IMU sensor and images of respective time frames. Some files had normal time frames only, while other files contained both normal and abnormal time frames which are mixed. The task was to find the abnormal time frames using unsupervised methods. Usage of only the normal data was allowed during training. This essentially means we are unaware of the exact type and level of abnormal situation beforehand. As a result, we had to perform training and other calculations solely on normal data and using it we had to find the abnormal cases. Total 12 Rosbag files were given where the total number of normal images is 277 and the number of mixed images is 392.

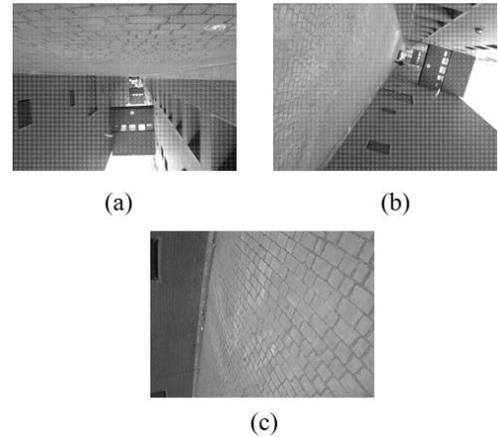

Figure 1: Example of image samples, (a) normal state, (b) abnormal state with the object and (c) abnormal state without the object

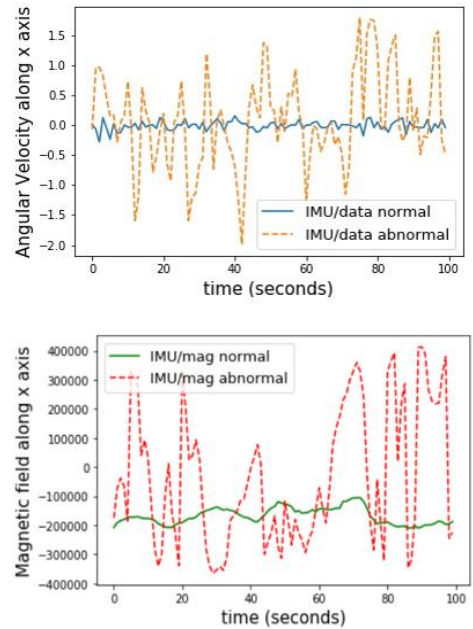

Figure 2: Normal and abnormal samples for 100 seconds from IMU/data and IMU/mag

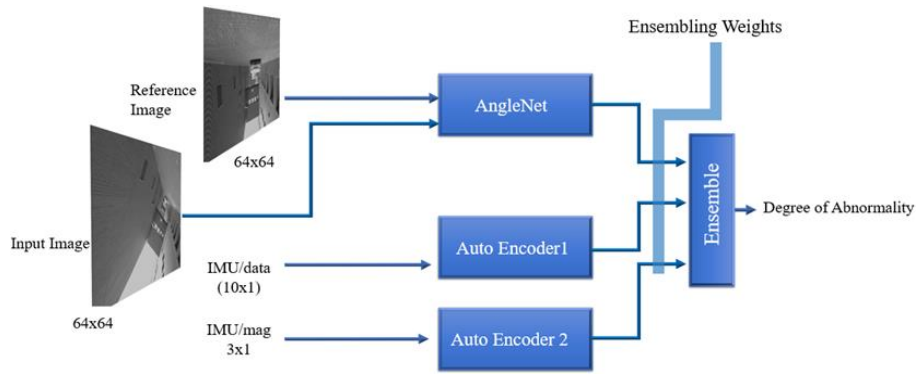

Figure 3: Schematic illustration of the proposed method for estimating degree of abnormality

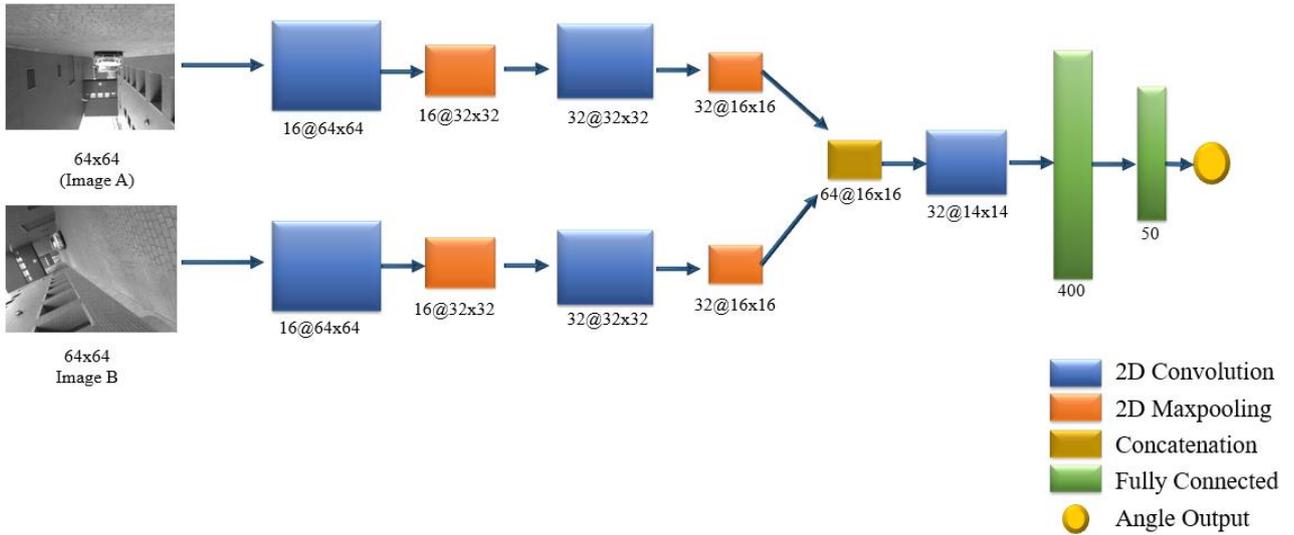

Figure 4: AngleNet used for estimation of angle difference between two images

Besides image data, IMU sensor data were provided. There are 6 types of data under IMU topicname among which we have used IMU/data and IMU/mag. A total of 987 normal timestamps was provided. IMU/data contains the orientation and velocity information of the drone along 3 axes and IMU/mag contains magnetic field data read by magnetometer. There are two separate parts in this detection procedure, image analysis and IMU sensor data analysis. We used an autoencoder based anomaly detection system for IMU data and used AngleNet to estimate the angle of an input image with respect to a normal reference image sample and later ensembled the 3 outputs to estimate the degree of abnormality. Figure 1 shows some normal and abnormal image samples provided in this dataset and it is very clear that the angle of view is changed significantly between the two cases. In this paper, we have considered two types of anomalies for images according to the figure. One is angle difference with normal image and the other is the absence of the object which is being followed by the drone. Moreover, to detect abnormal timestamps from IMU sensor data, there must be some distinctive factors between normal and abnormal IMU data which is demonstrated in Figure 2. Angular velocity and reading of magnetic field for 100 normal and abnormal samples along x-axis are presented in this figure. The variance of the data for the abnormal cases is quite clearly much larger in a time window compared to normal samples. These observations lead us to designing our proposed detection scheme which is described next.

### III. PROPOSED METHOD

This section describes the method we have used to develop an unsupervised model to detect abnormalities using the image and IMU sensor data. When the abnormal images were taken, the drone due to some malfunction became rotated at a significant angle. To accurately estimate the angle without depending on the abnormal data, we have introduced AngleNet which is used in an unsupervised manner. Besides we have used autoencoder based algorithm for modeling normal IMU data. Figure 3 demonstrates how we have combined the three outputs altogether to estimate the degree of anomaly. As per the figure, rather than using the outputs of the models directly, we have integrated weights manually and estimated the final degree of anomaly. The intuition is, AngleNet can only estimate the angle, not measure it precisely and for a single timestamp, the rotation of the drone may be slightly higher than the threshold angle but the sensor data may be significantly anomalous. Also, it can happen vice versa. So it is appropriate to use weights to combine them fruitfully and these are discussed further in Section IV. The different blocks of the proposed approach are described next in detail.

*A. AngleNet*

In the abnormal state of a surveillance drone, it is mostly the tilt angle that varies from the normal state as depicted in Fig. 2. In normal conditions, the drone is pretty stable as shown in the dataset. While for the unstable drone, the image is tilted at a significant angle. Inspired by Siamese network [21] we introduce AngleNet, a convolutional neural network based regression architecture to estimate significant angle change from the normal state. Previously Spyros *et al.* [22] introduced RotNet but we cannot use this model in this case as it works in a classification manner and can only detect angles among 0, 90, 180, 270 degrees. But in AngleNet, we tried to demonstrated angle estimation as a regression problem to have a more precise estimation of the degree of anomaly. In this model, a normal image should be provided first, and then the upcoming frames will be taken as input and the output is the angle between them. If there is a significant difference between the images such as object mismatch, the output will be significantly high which is why [23] or other classical computer vision-based angle estimation systems are not suitable for our application. In Figure 4, we see the model structure where it takes two images at a size of 64x64 pixels and after passing through several convolutional (Conv) blocks followed by maxpooling for each image branch, the activations are concatenated. This output is then processed by another Conv block and finally two fully connected hidden layers to produce the final angle estimation. On the layers, m@axb means a tensor having a depth of m channels and height and width of a rows and b columns respectively.

As discussed in section II, our goal is to design an unsupervised scenario, the images from the abnormal cases are not available during training. As a result, we have not trained AngleNet in a supervised manner using the specific application data. To train this model alternatively in this case, we have used the Stanford car dataset [24] to pre-train. The images were first augmented as so it can mimic the angle change, as demonstrated in Figure 5. The activation function of the final layer was ReLU, a non-negative linear function. Finally, there were 48,000 images which were divided into 80% train image and 20% validation image. Mean Squared Error was used as the loss function.

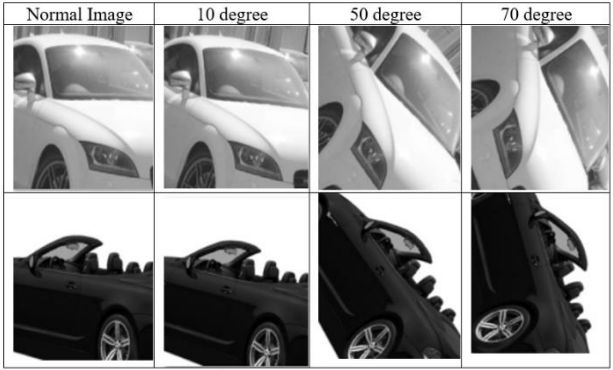

Figure 5: Samples of normal images and augmented images from Stanford Car Dataset [24]

After pre-training the model using the dataset, we have fine-tuned using the provided dataset [20] to estimate the degree of abnormality $\sigma_d$ of the images by dividing the output angle by 90.0 degrees. The performance of the model on classifying normal images is demonstrated in Section IV.

### B. Autoencoder based AD of IMU data

For both the IMU/data and IMU/mag, we have used autoencoder based anomaly detection system, similar to [25] but not exactly. The autoencoder models used here are shown in Figure 6. While training the autoencoders, it is supposed that in an abnormal time frame, both IMU/mag and IMU/data reading are abnormal. As there is no clear description of the dataset on this issue, we have considered as such and it produced a good result in our practical experimentation which is described in Section IV-F. As we are considering both are abnormal or normal at the same time frame, we have trained two autoencoders together. In this case, mean squared error (MSE) was considered as reconstruction loss. If L1 is the MSE for IMU/data sample and L2 is mse for IMU/mag sample, then:

$$L_1 = \frac{1}{n}\sum_{k=0}^{n}(y_k - \hat{y}_k)^2 \quad (1)$$

$$L_2 = \frac{1}{n}\sum_{i=0}^{n}(y_i - \hat{y}_i)^2 \quad (2)$$

where $y_k$ and $\hat{y}_k$ are ground truth and reconstructed output respectively for IMU/data samples. And $y_i$ and $\hat{y}_i$ are ground truth and reconstructed output for IMU/mag samples. The training loss is defined by the linear addition of the two losses, L1+L2. While training these two autoencoders, we have used normal samples only as we have discussed earlier. The results are discussed in Section IV.

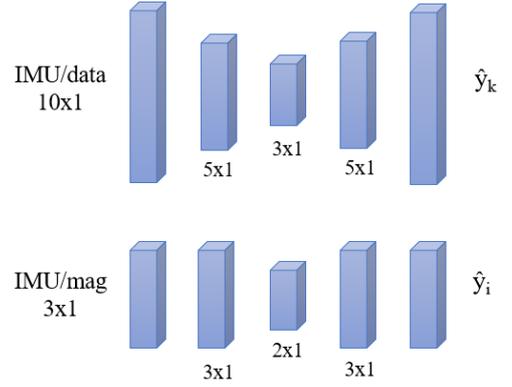

Figure 6: Autoencoder architecture used for IMU/data and IMU/mag

To find the degree of anomaly in this manner we had to consider $L_{max, data}$ and $L_{max, mag}$ as maximum reconstruction errors from the reconstruction errors of all individual samples from IMU/data and IMU/mag respectively. And calculated $\sigma_d$, $\sigma_m$ using (3) and (4).

$$\sigma_d = \frac{L_1}{L_{max,data}} \quad (3)$$

$$\sigma_m = \frac{L_1}{L_{max,mag}} \quad (4)$$

### C. Ensuring Adversarial Defense

As we are dealing with images delivered from the dataset, there is a possibility of being attacked by adversarial samples which will hamper the model to take decisions and bypass the pipeline. Although using original images as input has shown fair results, applying generated adversarial samples has caused significant performance degradation of our system which indicates the vulnerability that is further discussed in section IV-D.

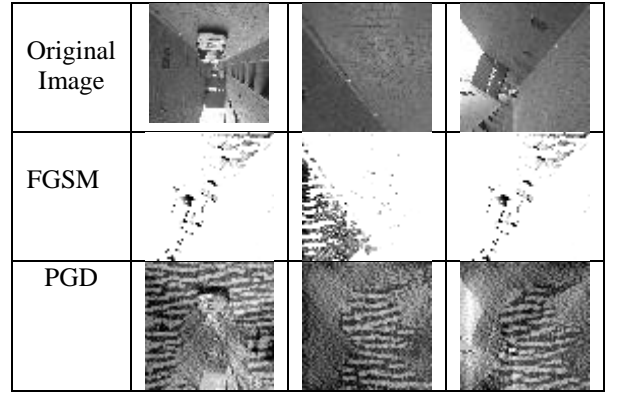

Figure 7: Original images and corresponding adversarial images

Then we perform adversarial training [26] as defense by training our model again with both the original images and generated adversarial images and achieved considerable accuracy. We have experimented using Projected Gradient Descent (PGD) and Fast-gradient-sign method (FGSM) attacks. For this kind of image, the PGD attack seems more suitable if we analyze the images of Figure 7.

## IV. EXPERIMENT AND RESULT

In this section, we have discussed the training procedures and results on the IEEE SP Cup 2020 dataset [20]. As this dataset is novel and contains two types of data, we have discussed the results separately and lastly, we have ensembled the results and produced the final output. Also, the performance of two types of adversarial attacks and defense are discussed.

### A. IMU Anomaly Detection

As discussed in Section II, we have used the autoencoder based anomaly detection system for IMU data. The results are discussed in Table 1. Normalization improves performance for IMU/mag as it has a wide range of data, from

TABLE 1. PERFORMANCE ON IMU DATA

| Data | Accuracy | F-1 Score | False Negative |
|---|---|---|---|
| IMU/data | 96.8% | 0.9812 | 2 |
| IMU/mag | 100% | 0.95 | 0 |

(-400000, 400000), it's very tough to converge the loss without normalization. As discussed in [27], the authors have claimed slightly better accuracy than us on the same dataset but their system is resources expensive. We are using two tiny autoencoders as shown in Figure 6 with a shared loss for IMU anomaly detection which can be run on common embedded devices of a low resource like Raspberry pi, which is validated in subsection *G*. Even though the accuracy reported in [27] excels ours, we can run our system in much lower resource-budget having very few or no false-negative cases. While accuracy is an important metric, we also need to compare false positive and false negatives to provide a fairer comparison. The F-1 score and false negative ratio is given in Table 1, which show satisfactory performance. The comparison among some other popular algorithms is given in Table 2. Notably, the proposed algorithm in this paper was among the top 10 performing submissions in IEEE SP Cup 2020.

TABLE 2: COMPARISONS WITH OTHER ALGORITHMS

| Algorithm | Accuracy |
|---|---|
| Autoencoder, Rad *et. al.* [28] | 96.87% |
| Kmeans Clustering, Iqbal *et. al.* [15] | 93.28% |
| 1D CNN, Kiranyaz *et. al.* [29] | 89.9% |
| PCA and Kmeans of IMU | 82.3% |
| Spectral Clustering of IMU | 91.7% |
| Clustering of IMU with image classification [30] | 97.3% |
| This paper | **97.8%** |

### B. AngleNet Based Anomaly Detection

Using AngleNet, we can estimate the angle between the test image and the normal image. In the abnormal images, the rotation angle is the main distinctive factor. Any images rotated by 30 degrees is supposed to be abnormal. But the threshold is perfectly tunable and user-defined. The performance of AngleNet on the test images is shown in Table 3.

TABLE 3. PERFORMANCE OF ANGLENET ON NORMAL IMAGE

| Threshold Angle | Accuracy |
|---|---|
| 30 | 94.7% |
| 20 | 86.4% |

The state-of-the-art image/video novelty detection algorithms is mostly for environmental anomaly detection which is not perfectly inclined with the problem we have worked with. We are more interested to find the anomaly of the device rather than the environment. Some comparisons with various methods are shown in Table 4.

TABLE 4: COMPARISON OF ANOMALY DETECTION USING IMAGE

| Algorithm | Accuracy |
|---|---|
| AngleNet | **94.7%** |
| Optical flow supervised | 89.4% |
| Optical flow, unsupervised | 92.33% |
| Binary Classification | 84.85% |
| Spectral Clustering | 91.7% |

The angle is not only the difference between normal and abnormal image samples but also there are some motion factors. So, we have compared the performance of anomaly detection between using optical flow of normal and actual images. In the dataset, the provided images were sampled so they did not have a gradual movement shift among them, rather there is a rapid difference in motion and content, so optical flow did not perform so well in this case.

### C. Ensembling Models

The process is designed so that both the clustering-based anomaly detection and AngleNet can be used separately or in an ensemble manner. As we have discussed, we calculate the degree of abnormality in each case, ensembling them can produce a combined result. Proposed ensembling formula is given as-

$$N = w_d * \sigma_d + w_m * \sigma_m + w_I * \sigma_I \qquad (5)$$

where N is the combined degree of abnormality, $w_d$, $w_m$, $w_I$ represents the weights for three different models such as two autoencoder-based models and one Convolutional Neural Network-based model, AngleNet. And $\sigma_d$, $\sigma_m$, $\sigma_I$ represent the degree of abnormality for IMU/data, IMU/mag, and Image respectively. After ensembling those models, we have achieved an overall 97.8% accuracy with an F-1 score of 0.98. To find the 3 weights, first, we considered 1 for each of them primarily and performed extensive experiments to empirically determine the values of the weights. Our investigation finds

$w_I = 0.75$, $w_m = 0.9$ and $w_d = 1$ provides the best results and these values were chosen for subsequent analysis. We considered a timestamp to be abnormal if $N \geq 1$.

*D. Adversarial Defense*

As discussed in Section III, we have used FGSM and PGD to generate adversarial samples for our model which can confuse the model. We used the Cleverhans library to generate the attacks. For FGSM, we used an $l_\infty$ attack with $\epsilon = 0.25$. For the PGD attack, a similar $\epsilon$ was used for 20 iterations. In both cases, images were clipped to [0, 1]. The adversarial attack accuracy is shown in Table 5. As indicated, our model has performed rather poorly on adversarial samples, especially for PGD. We have tested both optical flow and original images for generating adversarial samples. The intuition behind using optical flow was, such a scenario can easily be generated so the optical flow is similar to any adversarial optical flow, rather than generating original adversarial scenarios for the drone camera.

TABLE 5: PERFORMANCE ON ADVERSARIAL SAMPLES

| Input Type | Algorithm | Attack Accuracy | Defense Accuracy |
|---|---|---|---|
| Original Image | FGSM | 64% | 89.9% |
|  | PGD | 4% | 92.3% |
| Optical Flow | FGSM | 47.7% | 84.5% |
|  | PGD | 9.89% | 73.46% |

It's clear that our adversarial defense system is more vulnerable to optical flow inputs. Also, practically it is more difficult to create adversarial optical flow than adversarial input image samples. Interestingly, these adversarial perturbations are usually added after the clean image has been captured. Since our drone camera is taking the images and processing it on the fly, a natural query might be about the nature of successfully accomplishing such fishing attacks. In this respect, we consider two separate scenarios. The most prominent and direct form of attack could be physical attacks on the ML detection system. It has been shown that adversarial attack can be performed by placing intentional objects which fool the system [31] like adversarial turtle or using adversarial patches [32]. While it is difficult to model all possible cases regarding this type of attack, we generated 5 types of patches and trained our model putting these patches at various places around the perimeter that the drone is moving around. The results are shown in Table 6, row 1 which shown the training is quite successful in regaining the accuracy. The other form of attack could be possible through network intrusion if the classification is performed on the cloud. In this case, due to resource-constraints, it may not be possible to process the collected data on the autonomous agent itself, so data is transmitted to a remote server and detection is performed there. For such cases, if the adversary intrudes our network and corrupts the transmitted data, then it may cause wrong detections. Considering this scenario, there could be two possible cases. First is the data dependent adversarial perturbations are added each image or sensor data separately like FGSM or PGD attacks. This is the case considered as discussed above in Table 5. However, calculating such data-dependent adversarial noise takes time which in attack cases during real-time transmission may not be feasible, especially with no direct access to the data for the adversary. We believe a more realistic threat model in this case will be image-agnostic perturbations like the universal adversarial perturbations (UAP) [33]. These can be pre-computed and are common noise added to every image, as a result could potentially be added to the gathered data by the drone provided that the attacker has been able to perform network intrusion. To counter such probable attacks, we compute some UAPs for our training images (both normal as well as abnormal) and perform adversarial training by augmenting the dataset with these images. Again, the system is now able to withstand such attack quite well as depicted in row 2 of Table 6. Finally, another form of attack could be through modeling the distribution of the training data manifold and using generative models to create perturbations [34]. Although such attacks are quite unlikely to occur in real-life cases, as a precautionary measure, we also generate such images and augment our training. Notably, even if the adversaries are unlikely for this specific application, we anticipate such data augmentation to be helpful in better generalization of our model.

TABLE 6: PERFORMANCE AGAINST DIFFERENT ADVERSARIAL ATTACK SCHEMES

| Attack Scheme | Attack Accuracy | Defense Accuracy |
|---|---|---|
| Physical Attack [32] | 4% | 84.6% |
| UAP [33] | 24% | 91.5% |
| NAG [34] | 14.7% | 87.4% |

*E. Computational Cost*

While training AngleNet on the Car dataset, we have used GoogleColab with Nvidia Tesla K80 GPU with 12 Gigabytes of memory. But for the testing purpose, it runs on a computer with 2 Gigabytes of GPU seamlessly. The system is tested on a system containing the Intel Core i5 processor, 8 Gigabytes of RAM and Nvidia 940 MX. It takes 0.47 seconds on an average to process a single frame.

*F. Implementation on In-house setup*

For demonstration of real-time usage on an embedded device and verifying robustness, we deployed our proposed method on a raspberry pi where the autoencoders ran on the pi and CNN based processing works on a remote server. The system was tested on in-house setup, with custom hexacopter running on Ardupilot and we used raspberry pi 3 for real-time processing and sending video frames to the server. In this setup, the accuracy of the algorithm turned out to be 93.69%. This proves that the system is able to adapt to different test conditions and perform satisfactorily in real-time.

## V. CONCLUSION

In this paper, we have demonstrated an ensembled approach for unmanned vehicle anomaly detection. Our approach does not classify any sample strictly normal or abnormal, rather we have used the degree of abnormality as a metric for prediction. The lower the value, the closer it is to normal situation. We have introduced AngleNet which is used to estimate angles between two input images and using this angle we can determine anomaly. We have trained AngleNet on the Stanford Car Dataset and used transfer learning to estimate the angle of the images of the SP Cup dataset. For detecting abnormal IMU samples, we have used an autoencoder based anomaly detection system. Along with maintaining good accuracy, we have integrated defense mechanisms from adversarial attacks which is a potential threat to our algorithm. Again, this algorithm performs robustly in real-time for an in-house dataset as well. Future work includes incorporating both device and environmental anomaly detection in a single scheme for a comprehensive analysis of surveillance systems.

## REFERENCES

[1] S. S. Chowdhury, C. Lee & K. Roy, "Towards Understanding the Effect of Leak in Spiking Neural Networks," *arXiv preprint arXiv:2006.08761* (2020).

[2] R. Hyder, S. S. Chowdhury, and S. A. Fattah, "Real-time non-intrusive eye-gaze tracking based wheelchair control for the physically challenged," in 2016 IEEE EMBS Conference on Biomedical Engineering and Sciences (IECBES), pp. 784–787, IEEE, 2016.

[3] S. S. Chowdhury, R. Hyder, C. Shahanaz, and S. A. Fattah, "Robust single finger movement detection scheme for real time wheelchair control by physically challenged people," in 2017 IEEE Region 10 Humanitarian Technology Conference (R10-HTC), pp. 773–777, IEEE, 2017.

[4] A. Maksud, R. I. Chowdhury, T. T. Chowdhury, S. A. Fattah, C. Shahanaz, and S. S. Chowdhury, "Low-cost eeg based electric wheelchair with advanced control features," in TENCON 2017-2017 IEEE Region 10 Conference, pp. 2648–2653, IEEE, 2017.

[5] C. Shahnaz, A. Maksud, S. A. Fattah, and S. S. Chowdhury, "Lowcost smart electric wheelchair with destination mapping and intelligent control features," in 2017 IEEE International Symposium on Technology and Society (ISTAS), pp. 1–6, IEEE, 2017.

[6] S. A. Fattah, N. M. Rahman, A. Maksud, S. I. Foysal, R. I. Chowdhury, S. S. Chowdhury, and C. Shahanaz, "Stetho-phone: Low-cost digital stethoscope for remote personalized healthcare," in 2017 IEEE Global Humanitarian Technology Conference (GHTC), pp. 1–7, IEEE, 2017.

[7] S. S. Chowdhury, S. M. A. Uddin, and E. Kabir, "Numerical analysis of sensitivity enhancement of surface plasmon resonance biosensors using a mirrored bilayer structure," Photonics and Nanostructures-Fundamentals and Applications, p. 100815, 2020.

[8] S. M. A. Uddin, S. S. Chowdhury, and E. Kabir, "A theoretical model for determination of optimum metal thickness in kretschmann configuration based surface plasmon resonance biosensors," in 2017 International Conference on Electrical, Computer and Communication Engineering (ECCE), pp. 651–654, 2017.

[9] S. S. Chowdhury, R. Hyder, M. S. B. Hafiz, and M. A. Haque, "Real-time robust heart rate estimation from wrist-type ppg signals using multiple reference adaptive noise cancellation," IEEE journal of biomedical and health informatics, vol. 22, no. 2, pp. 450–459, 2016.

[10] S. S. Chowdhury, M. S. Hasan, and R. Sharmin, "Robust heart rate estimation from ppg signals with intense motion artifacts using cascade of adaptive filter and recurrent neural network," in TENCON 2019-2019 IEEE Region 10 Conference (TENCON), pp. 1952–1957, IEEE, 2019.

[11] M. Baydoun, D. Campo, V. Sanguineti, L. Marcenaro, A. Cavallaro, and C. Regazzoni, "Learning Switching Models for Abnormality Detection for Autonomous Driving," *2018 21st International Conference on Information Fusion (FUSION)*, 2018.

[12] Xiaoyong Yuan, Pan He, Qile Zhu, Xiaolin Li: "Adversarial Examples: Attacks and Defenses for Deep Learning", 2017; *arXiv:1712.07107*.

[13] D. Campo, M. Baydoun, P. Marin, D. Martin, L. Marcenaro, A. D. L. Escalera, and C. Regazzoni, "Learning Probabilistic Awareness Models for Detecting Abnormalities in Vehicle Motions," *IEEE Transactions on Intelligent Transportation Systems*, vol. 21, no. 3, pp. 1308–1320, 2020.

[14] D. Kanapram, P. Marin-Plaza, L. Marcenaro, D. Martin, A. D. L. Escalera, and C. Regazzoni, "Self-awareness in Intelligent Vehicles: Experience Based Abnormality Detection," *Advances in Intelligent Systems and Computing Robot 2019: Fourth Iberian Robotics Conference*, pp. 216–228, 2019.

[15] H. Iqbal, D. Campo, M. Baydoun, L. Marcenaro, D. M. Gomez, and C. Regazzoni, "Clustering Optimization for Abnormality Detection in Semi-Autonomous Systems," *1st International Workshop on Multimodal Understanding and Learning for Embodied Applications - MULEA '19*, 2019.

[16] M. Baydoun, M. Ravanbakhsh, D. Campo, P. Marin, D. Martin, L. Marcenaro, A. Cavallaro, and C. S. Regazzoni, "a Multi-Perspective Approach to Anomaly Detection for Self -Aware Embodied Agents," *2018 IEEE International Conference on Acoustics, Speech and Signal Processing (ICASSP)*, 2018.

[17] D. Sirkin, N. Martelaro, M. Johns, and W. Ju, "Toward Measurement of Situation Awareness in Autonomous Vehicles," *Proceedings of the 2017 CHI Conference on Human Factors in Computing Systems - CHI '17*, 2017.

[18] M. Ravanbakhsh, M. Baydoun, D. Campo, P. Marin, D. Martin, L. Marcenaro, and C. S. Regazzoni, "Hierarchy of Gans for Learning Embodied Self-Awareness Model," *2018 25th IEEE International Conference on Image Processing (ICIP)*, 2018.

[19] M. Baydoun, M. Ravanbakhsh, D. Campo, P. Marin, D. Martin, L. Marcenaro, A. Cavallaro, and C. S. Regazzoni, "a Multi-Perspective Approach to Anomaly Detection for Self -Aware Embodied Agents," *2018 IEEE International Conference on Acoustics, Speech and Signal Processing (ICASSP)*, 2018.

[20] "Unsupervised abnormality detection by using intelligent and heterogeneous autonomous systems,'" *[5] ICASSP*. [Online]. Available: https://2020.ieeeicassp.org/authors/sp-cup-2020/. [Accessed: 23-Mar-2020].

[21] Gregory Koch. Siamese neural networks for one-shot image recognition. PhD thesis, University of Toronto, 2015.

[22] Spyros Gidaris, Praveer Singh, Nikos Komodakis: "Unsupervised Representation Learning by Predicting Image Rotations", 2018; *arXiv:1803.07728*.

[23] R. Qian, W. Li and N. Yu, "High precision rotation angle estimation for rotated images," *2013 IEEE International Conference on Multimedia and Expo Workshops (ICMEW)*, San Jose, CA, 2013, pp. 1-4, doi: 10.1109/ICMEW.2013.6618298.

[24] "Cars Dataset - Stanford AI Lab." [Online]. Available: http://ai.stanford.edu/~jkrause/cars/car_dataset.html. [Accessed: 24-Mar-2020].

[25] A. Borghesi, A. Bartolini, M. Lombardi, M. Milano, and L. Benini, "Anomaly Detection Using Autoencoders in High Performance Computing Systems," *Proceedings of the AAAI Conference on Artificial Intelligence*, vol. 33, pp. 9428–9433, 2019.

[26] Aleksander Madry, Aleksandar Makelov, Ludwig Schmidt, Dimitris Tsipras, Adrian Vladu: "Towards Deep Learning Models Resistant to Adversarial Attacks", 2017; *arXiv:1706.06083*.

[27] Bahavan, Nadarasar & Suman, Navaratnarajah & Cader, Sulhi & Ranganayake, Ruwinda & Seneviratne, Damitha & Maddumage, Vinu & Seneviratne, Gershom & Supun, Yasinha & Wijesiri, Isuru & Dehigaspitiya, Suchitha & Tissera, Dumindu & Edussooriya, Chamira. (2020). Anomaly Detection using Deep Reconstruction and Forecasting for Autonomous Systems.

[28] Mohammadian Rad, Nastaran & van Laarhoven, Twan & Furlanello, Cesare & Marchiori, Elena. (2018). Novelty Detection using Deep Normative Modeling for IMU-Based Abnormal Movement Monitoring in Parkinson's Disease and Autism Spectrum Disorders. Sensors. 18. 3533. 10.3390/s18103533.

[29] Kiranyaz, Serkan & Zabihi, Morteza & Bahrami Rad, Ali & Tahir, Anas & Ince, Turker & Ridha, Hamila & Gabbouj, Moncef. (2019).

Real-time PCG Anomaly Detection by Adaptive 1D Convolutional Neural Networks.

[30] S. S. Chowdhury, K. M. Islam, and R. Noor, (2020). Unsupervised Abnormality Detection Using Heterogeneous Autonomous Systems. *arXiv preprint arXiv:2006.03733*.

[31] Kurakin, A., Goodfellow, I. and Bengio, S., 2016. Adversarial examples in the physical world. *arXiv preprint arXiv:1607.02533*.

[32] Brown, Tom B., Dandelion Mané, Aurko Roy, Martín Abadi, and Justin Gilmer. "Adversarial patch." *arXiv preprint arXiv:1712.09665* (2017).

[33] Moosavi-Dezfooli, Seyed-Mohsen, Alhussein Fawzi, Omar Fawzi, and Pascal Frossard. "Universal adversarial perturbations." In *Proceedings of the IEEE conference on computer vision and pattern recognition*, pp. 1765-1773. 2017.

[34] Reddy Mopuri, Konda, Utkarsh Ojha, Utsav Garg, and R. Venkatesh Babu. "NAG: Network for adversary generation." In *Proceedings of the IEEE Conference on Computer Vision and Pattern Recognition*, pp. 742-751. 2018.